\documentclass[prl,superscriptaddress,twocolumn,preprintnumbers]{revtex4}

\usepackage{amsfonts}
\usepackage{amsmath}
\usepackage{amssymb}
\usepackage{bm}
\usepackage{dcolumn}
\usepackage{epsfig}
\usepackage{graphicx}
\usepackage{graphics}
\usepackage[latin1]{inputenc}
\usepackage{latexsym}
\usepackage{rotating}
\usepackage{hyperref}

\newcommand\be{\begin{equation}}
\newcommand\ba{\begin{eqnarray}}
\newcommand\ee{\end{equation}}
\newcommand\ea{\end{eqnarray}}

\begin{document}
\title {A New PPN Parameter to Test Chern-Simons Gravity}

\author{Stephon Alexander}
\affiliation{Institute for Gravitational Physics and Geometry, Center
  for Gravitational Wave Physics and Department of Physics, The
  Pennsylvania State University, University Park, PA 16802, USA} 

\author{Nicolas Yunes}
\affiliation{Institute for Gravitational Physics and Geometry, Center
  for Gravitational Wave Physics and Department of Physics, The
  Pennsylvania State University, University Park, PA 16802, USA} 

\date{\today}

\preprint{IGPG-07/3-4}

\begin{abstract}
  
  We study Chern-Simons (CS) gravity in the parameterized
  post-Newtonian (PPN) framework through a weak-field solution of the
  modified field equations. We find that CS gravity possesses the same
  PPN parameters as general relativity, except for the inclusion of a
  new term, proportional to the CS coupling and the curl of the PPN
  vector potential. This new term leads to a modification of frame
  dragging and gyroscopic precession and we provide an estimate of its
  size. This correction might be used in experiments, such as Gravity
  Probe B, to bound CS gravity and test string theory.

\end{abstract}

\maketitle

{\emph{Introduction}}.  Current astronomical observations, such as the
apparent acceleration of the universe, suggest a possible infrared
modification to general relativity (GR). In the same spirit, another
unresolved problem of cosmology, the cosmic baryon asymmetry, suggests
a modification of general relativity via the inclusion of a
Chern-Simons (CS) correction during the inflationary
period~\cite{alexander:2004:lfg}. This Chern-Simons correction is not
an {\emph{ad-hoc}} extension, but it is actually motivated by both
string theory, as a necessary anomaly-canceling term to conserve
unitarity~\cite {Polchinski:1998rr}, and loop quantum
gravity~\cite{Ashtekar:1988sw}.  Recently, imprints of CS gravity have
been investigated in the gravitational wave spectrum of the Cosmic
Microwave Background (CMB), where it was found to produce a circular,
$V$-mode, polarization, albeit marginally
detectable~\cite{Lue:1998mq}.  Motivated by observational signatures
of string theory and loop-quantum gravity, we will explore and develop
a new observational window to distinguish CS gravity from classical
GR, which is of direct interest to gravitational experiments currently
underway, such as Gravity Probe B (GP B)~\cite{GravityProbeB} and
lunar ranging~\cite{Murphy:2007nt}.

A proven avenue for testing alternative theories of gravity with
current solar-system experiments is the parameterized post-Newtonian
(PPN) framework~\cite{Will:1993ns}. This framework considers
weak-field solutions of the field equations of the alternative theory
and expresses them in terms of PPN potentials and parameters. The PPN
potentials depend on the details of the system under consideration,
while the PPN parameters can be mapped to intrinsic parameters of the
theory. Predictions of the alternative theory can then be computed in
terms of PPN parameters and compared to solar-system experiments,
leading to stringent tests. One of the strengths of this framework is
its generality: a single super-metric with certain PPN parameters can
be constructed to reproduce and test several different alternative
theories~\cite{Will:1993ns} ({\emph{e.g.}}, scalar-tensor,
vector-tensor, bimetric and stratified theories.)  Other tests of
alternative theories of gravity have also been proposed, some of which
require a gravitational wave detection and shall not be discussed
here~\cite{will:1998:bmo,will:2004:tat,berti:2005:esb}.

In this letter, we present a parametrized PPN expansion of CS gravity
to allow for tests with current solar-system experiments. We discover
that CS gravity {\emph{demands}} the introduction of only one new term
to the PPN super-metric and, thus, one new PPN parameter. This new
term depends both on an intrinsic parameter of CS gravity, as well as
on the curl of the PPN vector potential. Such a coupling of CS gravity
to gravitational vector currents had so far been neglected.
Furthermore, curl terms in the super-metric had also been neglected by
the PPN community because other alternative theories had not required
them. We find that this new term captures the key physical effect of
CS gravity in the weak-field limit, leading to a modification of
frame-dragging that could be used to test this GR extension with GP
B~\cite{GravityProbeB}.

{\emph{CS Gravity in a Nutshell}}. CS gravity modifies GR via the
addition of a new term to the action, namely~\cite{jackiw:2003:cmo,Guarrera:2007tu}
\begin{equation}
\label{CS-action}
S_{CS} = \frac{1}{16 \pi G} \int d^4 x \frac{1}{4} f \; 
R ^{\star} R,
\end{equation}
where $G$ is the Newton's gravitational constant, $f$ is a prescribed
external quantity~\footnote{This field can also be interpreted as a
  dynamical one, in which case it possesses equations of motion.
  These equations dynamically enforce the Pontryagin
  constraint~\cite{Guarrera:2007tu}} (with units of squared length in
geometrized units) that acts as a coupling constant, $R$ is the Ricci
scalar and the star stands for the dual operation. The modified field
equations can be obtained by varying the action with respect to the
metric.  These equations, in trace-reversed form, are
\be
\label{mod-field-eqs}
R_{\mu \nu} + C_{\mu \nu} = 8 \pi \left(T_{\mu \nu} - \frac{1}{2}
  g_{\mu \nu} T \right),
\ee
where $C_{\mu \nu}$ is a Cotton-like tensor, $R_{\mu \nu}$ is the
Ricci tensor, $T_{\mu \nu}$ is a stress-energy tensor, with $T$ its
$4$-dimensional trace, and Greek letters range over spacetime indices.
The Cotton tensor encodes the CS modification to GR:
\be
C_{\mu \nu} = - \frac{1}{\sqrt{-g}} \left[ f_{,\sigma}
  \epsilon^{\sigma \alpha \beta}{}_{(\mu} D_{\alpha}
R_{\nu) \beta} + \left(D_{\sigma}f_{,\tau}\right) \;
{}^{\star}R^{\tau}{}_{(\mu}{}^{\sigma}{}_{\nu)} \right],  
\ee
where parenthesis stand for symmetrization, $g$ is the determinant of
the metric, $\epsilon^{\sigma \alpha \beta \mu}$ is the Levi-Civita
symbol~\footnote{We here correct a mistake in~\cite{Alexander:2007vt},
  where the Levi-Civita tensor was used instead of the symbol, which
  in the linearized theory are equivalent up to terms of
  ${\cal{O}}(h)^2$.}, $D_{\alpha}$ and colon subscripts stand for
covariant and partial differentiation respectively.

The CS correction to the action has been shown to lead to
birefringence in the polarization of gravitational
waves~\cite{alexander:2005:bgw}. In this context, birefringence is a
change in the amplitude of different polarization modes as the wave
propagates.  Recently, there have been
proposals~\cite{2001PhRvD..64h3007G} of astrophysical tests of
theories where gravitational waves with different polarization
propagate at different speeds, but this is not the case in CS gravity.
Nonetheless, such amplitude birefringence in gravitational waves could
have a signature in the anisotropies of the
CMB~\cite{alexander:2004:lfg} and could explain baryogenesis during
the inflationary epoch~\cite{Lue:1998mq}. Given that the CS extension
has been key in proposing a plausible explanation to some important
cosmological problems, it seems natural to study CS gravity in the
light of solar-system experiments. 

Can we understand the CS correction in more physical terms? For this
purpose, let us consider the CS coupling parameter $f$ as a
consequence of some external field that permeates all of spacetime,
such as a model-independent gravitational axion. This field could
depend on some intrinsic properties of spacetime, such as the
fundamental string scale~\cite{Brandenberger:1988:sit} or the
existence of warped compactifications~\cite{randall:1999:atc}.
Furthermore, this field could also be coupled to regions of high
curvature, such as binary neutron star systems, through standard
model-like currents. These couplings have been proposed as
enhancements to the CS modification, which would otherwise be
suppressed by the Planck scale. For simplicity, in this letter we
shall concentrate on a CS coupling parameter that is spatially
isotropic and whose only non-vanishing derivative is $\dot{f}$. These
assumptions are made such that time-translation symmetry and
reparameterization invariance are preserved in the modified
theory~\cite{jackiw:2003:cmo}.

{\emph{Weak Field Expansion of CS Gravity}}. Let us consider a system
that is weakly gravitating, such that we can expand the metric about a
fixed Minkowski background $\eta_{\mu \nu}$. In other words, let us
write $g_{\mu \nu} = \eta_{\mu \nu} + h_{\mu \nu}$, with $h_{\mu \nu}$
a small perturbation, and expand the Cotton tensor to second order in
$h_{\mu \nu}$. We then obtain a complicated expression that can is
schematically given by~\footnote{The full expansion of the Cotton
  tensor to second order in the metric perturbation is given
  in~\cite{Alexander:2007vt}}
\be
\label{Ricci-2nd-O}
C_{\mu \nu} \sim N^{(1)}_{\mu \nu}[{\epsilon} \cdot h'''] +
  N^{(2)}_{\mu \nu}[{\epsilon} \cdot h h'''] + N^{(3)}_{\mu
  \nu}[{\epsilon} \cdot h' h''], 
\ee
where primes stand for spatial or temporal derivatives and ${\epsilon}
\cdot A$ is the full contraction of the Levi-Civita symbol with the
tensor $A_{\mu_1 \ldots \mu_n}$. Note that here we have not assumed
any gauge conditions and, thus, Eq.~(\ref{Ricci-2nd-O}) could be used
in future work to calculate gravitational wave solutions to
${\cal{O}}(h)^2$.  Equation (\ref{Ricci-2nd-O}) to linear
order and in the Lorenz gauge [$h_{\mu \alpha,}{}^{\alpha} =
h_{,\mu}/2$, $h\equiv\eta^{\mu \nu} h_{\mu \nu}$] reduces to the
previously known expression~\cite{jackiw:2003:cmo}
\ba
\label{linear-Lorentz-filed-eqs}
C_{\mu \nu} &=& - \frac{\dot{f}}{2} {\epsilon}^{0 \alpha \beta}{}_{(\mu}
\square_{\eta} h_{\nu) \beta,\alpha} + {\cal{O}}(h)^2,
\ea
where $\eta_{\mu \nu}$ is the Minkowski metric and $\square_{\eta}$ is
the D'Alambertian associated with it.

Before proceeding with the PPN solution of the modified field
equations, we must discuss the stress-energy source that we shall
employ. Here we model this tensor as that of a perfect fluid
(cf.~eg.~\cite{Will:1993ns}). Such a stress-energy tensor is
sufficient to obtain the PPN solution of the modified field equations
for solar-system experiments, where the internal structure of the
fluid bodies shall be neglected to lowest order by the effacement
principle~\cite{Blanchet:2002av}.

The stress energy considered here requires the strong equivalence
principle (SEP) to hold in CS gravity~\cite{alexander:2004:lfg}.  This
principle states that the outcome of all local gravitational
experiments is independent of the experimenters reference frame. In
other words, the motion of test particles is exclusively governed by
the spacetime metric, through the divergence of the stress-energy
tensor. In CS gravity, the divergence of Eq.~(\ref{mod-field-eqs})
leads to
\be
\label{DivT}
D^{\alpha} C_{\alpha \beta} = \frac{\dot{f} \delta^{\beta 0}}{8
  \sqrt{-g}}  R^{\star} R = 8 \pi D^{\alpha} T_{\alpha \beta},
\ee
where $D^{\alpha} G_{\alpha \beta} = 0$ by the Bianchi identities. The
right-hand side of Eq.~(\ref{DivT}) vanishes in vacuum and, thus, the
SEP holds provided $R^{\star}R = 0$, which is known as the Pontryagin
constraint. In fact, in the original formulation of CS
gravity~\cite{jackiw:2003:cmo}, this constraint was independently
required to preserve time-translation symmetry and spatial
reparameterization invariance.  We shall later see that the solution
to Eq.~(\ref{mod-field-eqs}) found here automatically satisfies this
constraint to ${\cal{O}}(h)^2$ and, thus, the SEP holds.

{\emph{Weak Field Solution}}. Let us first study the weak-field
solution to the modified field equations in Lorenz gauge. The formal
first-order solution of Eq.~(\ref{mod-field-eqs}), with
Eq.~(\ref{linear-Lorentz-filed-eqs}) used for the Cotton tensor, is
simply~\cite{Alexander:2007vt}
\be
h_{\mu \nu} = -16 \pi \; \square_{\eta}^{-1} \left[\bar{T}_{\mu \nu} 
- \dot{f} \epsilon^{k \ell i} \left(\delta_{i (\mu} T_{\nu) \ell,k} -
   \frac{1}{2} \delta_{i (\mu} \eta_{\nu) \ell} T_{,k} \right) \right],
\ee
where $\bar{T}_{\mu \nu}$ is the trace reversed $T_{\mu \nu}$. Note
that this formal solution has the property that as $\dot{f} \to 0$ it
reduces to that predicted by the post-Newtonian (PN) expansion of
GR~\cite{Blanchet:2002av}. In fact, such a solution is the cornerstone
of the PN formalism and would be essential if one were to pursue such
an expansion of CS gravity.

Let us now proceed with the PPN solution of the modified field
equations. The PPN formalism differs from the PN Lagrangian
formulation for inspiraling compact binaries~\cite{Blanchet:2002av} by
the use of a different gauge, the harmonic one~\footnote{This gauge is
  defined by $\partial_{\mu} \widetilde{h}^{\alpha \mu} = 0$, with the
  metric perturbation defined via $\widetilde{h}^{\alpha \beta} =
  \eta^{\alpha \beta} - (-g)^{1/2} g^{\alpha \beta}$.}. In the PPN
formalism, one usually employs a PPN gauge, designed such that the
spatial part of the metric is diagonal and isotropic. These conditions
can be enforced perturbatively via~\cite{Will:1993ns}
\ba
\label{gauge}
h_{jk,}{}^{k} - \frac{1}{2} h_{,j} &=& {\cal{O}}(4),
\quad 
h_{0k,}{}^{k} - \frac{1}{2} h^{k}{}_{k,0} = {\cal{O}}(5), \quad
\ea
where $h^{k}{}_{k}$ is the spatial trace of the metric perturbation
and the symbol ${\cal{O}}(A)$ stands for PN remainders of order
${\cal{O}}(1/c)^A$, with $c$ the speed of
light~\cite{Alexander:2007vt}. One can show that Eq.~(\ref{gauge}) is
related to the Lorenz gauge via an infinitesimal gauge transformation.
The solution to the CS modified field equations in PPN gauge is given
by
\ba
\label{CS-full-metric}
g_{00} &=& -1 + 2 U - 2 U^2 + 4 \Phi_1 + 4 \Phi_2 + 2 \Phi_3 + 6 \Phi_4 +
{\cal{O}}(6), 
\nonumber \\
g_{0i} &=& -\frac{7}{2} V_i - \frac{1}{2} W_i + 2 \dot{f} \left(\nabla
  \times V\right)_i + {\cal{O}}(5),
\nonumber \\
g_{ij} &=& \left(1 + 2 U \right) \delta_{ij} + {\cal{O}}(4),
\ea
where $\{U,\Phi_1,\Phi_2,\Phi_3,\Phi_4,V_i,W_i\}$ are PPN potentials
(see, {\emph{e.g.}},~\cite{Will:1993ns} for definitions and discussion
of these potentials.) Both the PPN potentials and parameters take the
same values in CS gravity as in GR.  Equation (\ref{CS-full-metric})
is a solution to $1$ PN order, since from it one could calculate the
point-particle Lagrangian to ${\cal{O}}(4)$. As one can check, this
solution satisfies the Pontryagin constraint~\cite{Alexander:2007vt}.

Chern-Simons gravity introduces a correction to the metric in the
vectorial sector of the metric perturbation. This correction is
proportional to the first time derivative of the CS coupling
parameter, $\dot{f}$ and to the curl of the PPN vector potential
$V_i$. In principle, there is also a CS coupling to the other PPN
vector potential $W_i$, but this contribution is already accounted for
because $\nabla \times W_i = \nabla \times V_i$. Since this is the
only modification to the metric, the PPN parameters of CS gravity are
identical to those of classical GR, with the exception of the
inclusion of a new term in $g_{0i}$. In fact, defining the CS
correction as $\delta g_{0i} = g_{0i} - g_{0i}^{\mathrm{GR}}$, with
$g_{0i}^{\mathrm{GR}}$ the GR prediction, we get
\be
\label{CS-metric-mod}
\delta g_{0i} = \chi M \left(\nabla \times V\right)_{i},
\ee
where we have defined a new PPN parameter, $\chi \equiv = 2
\dot{f}/M$, with $M$ the characteristic mass scale of the source
inducing the vector potential. This new PPN parameter is rescaled by
$M$ to make it dimensionless and coordinate independent.  The
rescaling choice might seem arbitrary, but since $\dot{f}$ has units
of mass it can be interpreted as some CS mass scale, yielding $\chi
\propto m_{CS}/M$ as a ratio of masses with a clear physical meaning.

Until now, a PPN potential of the type of Eq.~(\ref{CS-metric-mod})
had not been considered, nor had any experimental constraints been
placed on $\chi$. Clearly, any experiment that samples the vectorial
sector of the metric perturbation, and thus, the frame-dragging
effect, could achieve such a constraint.

{\emph{Astrophysical Tests}}. Consider a system of $A$ nearly
spherical bodies in the standard PPN point-particle approximation,
where the PPN vector potential is~\cite{Will:1993ns}
\ba
\label{vector-pot}
V^i &=& \sum_A \frac{m_A}{r_A} v^i_A + \frac{1}{2} \sum_A
\left(\frac{J_A}{r_A^2} \times n_A\right)^i,
\ea
with $m_A$ the mass of the $A$th body, $r_A$ the field point distance
to the $A$th body, $n_A^i = x^i_A/r_A$ a unit vector pointing to the
$A$th body, $v_A$ the velocity of the $A$th body and $J_A^i$ the
spin-angular momentum of the $A$th body. When the number of bodies $A
= 2$, Eq.~(\ref{vector-pot}) is the vector potential for a binary of
spinning compact objects, while when there is only one body present
$A=1$ it represents the potential outside a moving spinning body. For
such a vector potential, the CS correction to the metric becomes
\be
\label{CS-term}
\delta g_{0i} = 2  \sum_A \frac{\dot{f}}{r_A} \left[ \frac{m_A}{r_A}
  \left(v_A \times n_A \right)^i - \frac{J^i_A}{2 r_A^2} + \frac{3}{2}
  \frac{\left(J_A \cdot n_A\right)}{r_A^2} n_A^i \right],
\ee
where the $\cdot$ and $\times$ operators are the flat-space inner and
cross products. Note that the CS correction couples both to the spin
and orbital angular momentum of the system.

The full gravitomagnetic sector of the metric becomes
\ba
\label{effective}
g_{0i} &=& \sum_A \left[ - \frac{7}{2} \frac{m_A}{r_A} v^i_A 
- \frac{m_A}{6 r_A} \left(v_A - v_A^{(eff)}\right)^i
\right.
\\ \nonumber 
&-& \left. 
\frac{1}{2} n_A^i \frac{m_A}{r_A} \left(v^{(eff)}_A \cdot n_A\right)
- 2  \left(\frac{J_A^{(eff)}}{r_A^2} \times n_A \right)^i \right],
\ea
where we have introduced an effective velocity and angular momentum
through
\ba
v_{A(eff)}^i &=& v^i_A - 6 \dot{f} \frac{J^i_A}{m_A r_A^2},
\quad
\label{J-eff}
J_{A(eff)}^i = J^i_A - \dot{f} m_A v_A^i,
\nonumber \\
\ea
When the spin angular momentum $J_A$ vanishes, $g_{0i}$ is identical
to that of a spinning moving object, with the spin induced by the CS
coupling to the orbital angular momentum. Such a coupling leads to an
interesting physical interpretation: if we model the field that
sources the CS coupling as a fluid that permeates all spacetime, the
CS modification to the metric is nothing but the ``dragging'' of such
a fluid~\cite{Alexander:2007vt}, whose strength is proportional to the
first derivative of the CS coupling parameter.

The CS correction to the metric computed here couples to the
three-velocity of the sources, which could suggest the possibility
that this effect is not coordinate-invariant. However, this
velocity-dependance comes directly from the PPN vector potential $V^i$
[Eq.~(\ref{vector-pot})]. Therefore, the CS correction to the metric
is as coordinate-invariant as the GR PPN metric itself, since this
also depends on $V^i$ [Eq.~(\ref{CS-full-metric})]~\cite{Will:1993ns}.
Observables, on the other hand, must be constructed in a
coordinate-invariant way, which is sensitive to the choice of basis
vectors. This choice in general depends on the experiments that look
for such observables~\cite{Will:2002ma} and a more formal analysis of
such coordinate issues should be carried out elsewhere.

We can now compute the correction to the frame-dragging effect in CS
gravity and compare it to the Lense-Thirring effect. Consider then a
free gyroscope in the presence of the gravitational field of
Eq.~(\ref{effective}). The gyroscope will acquire the precessional
$\Omega^i = (\nabla \times g)^i$, where $g^i = g_{0i}$. Therefore, the
CS modification to the precession angular velocity, defined via
$\delta \Omega^i = \Omega^i - \Omega^I_{GR}$, where $\Omega^i_{GR}$ is
the GR prediction, is given by
\be
\label{LT-correction}
\delta\Omega^i = - \sum_A \dot{f} \frac{m_A}{r_A^3} \left[ 3 \left(v_A \cdot
    n_A\right) n_A^i - v_A^i \right],
\ee
while the full Lense-Thirring term is
\be 
\label{full-ang}
\Omega^i_{LT} = - \frac{1}{r_A^3} \sum_A J_{A(eff)}^i - 3 n^i_A
\left(J_{A(eff)} \cdot n_A \right)^i,
\ee
which vanishes for static sources. As before, the CS correction has
the effect of modifying the classical GR prediction via the
replacement $J_A^i \to J^i_{A(eff)}$. 

If experiments~\cite{GravityProbeB} detect frame-dragging and find it
in agreement with the GR prediction, we can immediately test CS
gravity. In order to place a bound, however, a careful analysis must
be performed, using the tools developed in~\cite{Will:2002ma} and
properly accounting for the experiment's frame. Nonetheless, we can
construct a crude, order-of-magnitude estimate of the size of such a
bound. In order to do so, we assume the Newtonian limit ${\cal{O}}(J)
\sim {\cal{O}}(m R v)$, with $m$ the total mass of the system, $R$ the
distance from the gyroscope to the gravitational source (for GP B, $R
\sim 7000$ km~\cite{Will:2002ma}) and $v$ the orbital velocity, such
that we can model the CS correction as $|\Omega| \sim \Omega^{GR} (1 +
\dot{f}/R)$~\footnote{This can be seen by inserting ${\cal{O}}(J) \sim
  {\cal{O}}(m R v)$ into Eq.~(\ref{full-ang}), with the definition of
  the effective angular velocity given in Eq.~(\ref{J-eff}), and
  factoring out $1 + \dot{f}/R$.}.  Then, a one percent accuracy in
the frame-dragging measurement relative to the GR prediction (nominal
for GP B~\cite{Will:2002ma}) translates, roughly, into the bound
$\dot{f} \lesssim 10^{-3}$ seconds.

Let us conclude with a discussion of the scaling of the order of
magnitude of the CS correction. From Eqs.~(\ref{effective})
and~(\ref{LT-correction}), we can see that the CS correction is of
${\cal{O}}(3)$ if $\dot{f}/r_A$ is of order unity, which implies that
it is enhanced in regions of high-curvature, precisely where the PPN
and post-Newtonian analysis does not hold. Such scaling also suggests
that the CS effect might be larger in highly dynamical systems that
are not weakly-gravitating, such as compact object binaries. If
frame-dragging were measured in such systems to sufficient accuracy,
then possibly a much better bound could be placed on CS gravity.

{\emph{Conclusions}}. We have calculated the weak-field expansion of
CS gravity and solved the field equations in the PPN formalism. We
have found that CS gravity has the same PPN parameters as GR, except
for the inclusion of a new term in $g_{0i}$, which can be
parameterized in terms of a new PPN quantity. We have seen that this
new term leads to a correction to the frame-dragging effect, thus
allowing for the first solar-system test of CS gravity.

The CS correction is clearly enhanced in the non-linear regime, where
the stress-energy tensor diverges. This regime, however, is precisely
where the PN approximation and PPN framework break down. Therefore, an
accurate analysis of the size of the CS correction relative to the GR
prediction in the non-linear regime will have to await full numerical
simulations of modified GR. 

{\emph{Acknowledgements}}. We wish to thank Clifford Will, Roman
Jackiw, Pablo Laguna, and Ben Owen for insightful discussions. This
work was supported by NSF grants PHY-05-55-628 and the Center for
Gravitational Wave Physics, funded by the NSF via PHY-01-14375.

{\bf{Note added after submission:}} After submission of this work, a
paper was submitted that expands on the analysis presented here to
account for extended sources~\cite{Smith:2007jm}. After a detailed
study, a bound is placed on the CS coupling parameter with data from
LAGEOS and GP B.



\begin{thebibliography}{18}
\expandafter\ifx\csname natexlab\endcsname\relax\def\natexlab#1{#1}\fi
\expandafter\ifx\csname bibnamefont\endcsname\relax
  \def\bibnamefont#1{#1}\fi
\expandafter\ifx\csname bibfnamefont\endcsname\relax
  \def\bibfnamefont#1{#1}\fi
\expandafter\ifx\csname citenamefont\endcsname\relax
  \def\citenamefont#1{#1}\fi
\expandafter\ifx\csname url\endcsname\relax
  \def\url#1{\texttt{#1}}\fi
\expandafter\ifx\csname urlprefix\endcsname\relax\def\urlprefix{URL }\fi
\providecommand{\bibinfo}[2]{#2}
\providecommand{\eprint}[2][]{\url{#2}}

\bibitem[{\citenamefont{Alexander et~al.}(2006)\citenamefont{Alexander, Peskin,
  and Sheik-Jabbari}}]{alexander:2004:lfg}
\bibinfo{author}{\bibfnamefont{S.~H.~S.} \bibnamefont{Alexander}},
  \bibinfo{author}{\bibfnamefont{M.~E.} \bibnamefont{Peskin}},
  \bibnamefont{and} \bibinfo{author}{\bibfnamefont{M.~M.}
  \bibnamefont{Sheik-Jabbari}}, \bibinfo{journal}{Phys. Rev. Lett.}
  \textbf{\bibinfo{volume}{96}}, \bibinfo{pages}{081301}
  (\bibinfo{year}{2006}).

\bibitem[{\citenamefont{Polchinski}(1998)}]{Polchinski:1998rr}
\bibinfo{author}{\bibfnamefont{J.}~\bibnamefont{Polchinski}},
  \emph{\bibinfo{title}{String theory. Vol. 2: Superstring theory and beyond}}
  (\bibinfo{publisher}{Cambridge University Press},
  \bibinfo{address}{Cambridge, UK}, \bibinfo{year}{1998}).

\bibitem[{\citenamefont{Ashtekar et~al.}(1989)\citenamefont{Ashtekar,
  Balachandran, and Jo}}]{Ashtekar:1988sw}
\bibinfo{author}{\bibfnamefont{A.}~\bibnamefont{Ashtekar}},
  \bibinfo{author}{\bibfnamefont{A.~P.} \bibnamefont{Balachandran}},
  \bibnamefont{and} \bibinfo{author}{\bibfnamefont{S.}~\bibnamefont{Jo}},
  \bibinfo{journal}{Int. J. Mod. Phys.} \textbf{\bibinfo{volume}{A4}},
  \bibinfo{pages}{1493} (\bibinfo{year}{1989}).

\bibitem[{\citenamefont{Lue et~al.}(1999)\citenamefont{Lue, Wang, and
  Kamionkowski}}]{Lue:1998mq}
\bibinfo{author}{\bibfnamefont{A.}~\bibnamefont{Lue}},
  \bibinfo{author}{\bibfnamefont{L.-M.} \bibnamefont{Wang}}, \bibnamefont{and}
  \bibinfo{author}{\bibfnamefont{M.}~\bibnamefont{Kamionkowski}},
  \bibinfo{journal}{Phys. Rev. Lett.} \textbf{\bibinfo{volume}{83}},
  \bibinfo{pages}{1506} (\bibinfo{year}{1999}), \eprint{astro-ph/9812088}.

\bibitem[{Gra()}]{GravityProbeB}
\bibinfo{note}{A discussion of the history, technology and physics of Gravity
  Probe B can be found at http://einstein.standfod.edu}.

\bibitem[{\citenamefont{Murphy et~al.}(2007)\citenamefont{Murphy, Nordtvedt,
  and Turyshev}}]{Murphy:2007nt}
\bibinfo{author}{\bibfnamefont{J.}~\bibnamefont{Murphy}, \bibfnamefont{T.~W.}},
  \bibinfo{author}{\bibfnamefont{K.}~\bibnamefont{Nordtvedt}},
  \bibnamefont{and} \bibinfo{author}{\bibfnamefont{S.~G.}
  \bibnamefont{Turyshev}}, \bibinfo{journal}{Phys. Rev. Lett.}
  \textbf{\bibinfo{volume}{98}}, \bibinfo{pages}{071102}
  (\bibinfo{year}{2007}), \eprint{gr-qc/0702028}.

\bibitem[{\citenamefont{Will}(1993)}]{Will:1993ns}
\bibinfo{author}{\bibfnamefont{C.~M.} \bibnamefont{Will}},
  \emph{\bibinfo{title}{Theory and experiment in gravitational physics}}
  (\bibinfo{publisher}{Cambridge Univ. Press}, \bibinfo{address}{Cambridge,
  UK}, \bibinfo{year}{1993}).

\bibitem[{\citenamefont{Will}(1998)}]{will:1998:bmo}
\bibinfo{author}{\bibfnamefont{C.~M.} \bibnamefont{Will}},
  \bibinfo{journal}{Phys. Rev. D} \textbf{\bibinfo{volume}{57}},
  \bibinfo{pages}{2061} (\bibinfo{year}{1998}), \eprint{gr-qc/9709011}.

\bibitem[{\citenamefont{Will and Yunes}(2004)}]{will:2004:tat}
\bibinfo{author}{\bibfnamefont{C.~M.} \bibnamefont{Will}} \bibnamefont{and}
  \bibinfo{author}{\bibfnamefont{N.}~\bibnamefont{Yunes}},
  \bibinfo{journal}{Class. Quantum Grav.} \textbf{\bibinfo{volume}{21}},
  \bibinfo{pages}{4367} (\bibinfo{year}{2004}).

\bibitem[{\citenamefont{{Berti} et~al.}(2005)\citenamefont{{Berti}, {Buonanno},
  and {Will}}}]{berti:2005:esb}
\bibinfo{author}{\bibfnamefont{E.}~\bibnamefont{{Berti}}},
  \bibinfo{author}{\bibfnamefont{A.}~\bibnamefont{{Buonanno}}},
  \bibnamefont{and} \bibinfo{author}{\bibfnamefont{C.~M.}
  \bibnamefont{{Will}}}, \bibinfo{journal}{Phys. Rev. D}
  \textbf{\bibinfo{volume}{71}}, \bibinfo{pages}{084025}
  (\bibinfo{year}{2005}).

\bibitem[{\citenamefont{Jackiw and Pi}(2003)}]{jackiw:2003:cmo}
\bibinfo{author}{\bibfnamefont{R.}~\bibnamefont{Jackiw}} \bibnamefont{and}
  \bibinfo{author}{\bibfnamefont{S.~Y.} \bibnamefont{Pi}},
  \bibinfo{journal}{Phys. Rev.} \textbf{\bibinfo{volume}{D68}},
  \bibinfo{pages}{104012} (\bibinfo{year}{2003}).

\bibitem[{\citenamefont{Guarrera and Hariton}(2007)}]{Guarrera:2007tu}
\bibinfo{author}{\bibfnamefont{D.}~\bibnamefont{Guarrera}} \bibnamefont{and}
  \bibinfo{author}{\bibfnamefont{A.~J.} \bibnamefont{Hariton}}
  (\bibinfo{year}{2007}), \eprint{gr-qc/0702029}.

\bibitem[{\citenamefont{Alexander and Martin}(2005)}]{alexander:2005:bgw}
\bibinfo{author}{\bibfnamefont{S.}~\bibnamefont{Alexander}} \bibnamefont{and}
  \bibinfo{author}{\bibfnamefont{J.}~\bibnamefont{Martin}},
  \bibinfo{journal}{Phys. Rev.} \textbf{\bibinfo{volume}{D71}},
  \bibinfo{pages}{063526} (\bibinfo{year}{2005}), \eprint{hep-th/0410230}.

\bibitem[{\citenamefont{{Gleiser} and {Kozameh}}(2001)}]{2001PhRvD..64h3007G}
\bibinfo{author}{\bibfnamefont{R.~J.} \bibnamefont{{Gleiser}}}
  \bibnamefont{and} \bibinfo{author}{\bibfnamefont{C.~N.}
  \bibnamefont{{Kozameh}}}, \bibinfo{journal}{\prd}
  \textbf{\bibinfo{volume}{64}}, \bibinfo{pages}{083007}
  (\bibinfo{year}{2001}), \eprint{gr-qc/0102093}.

\bibitem[{\citenamefont{Brandenberger and Vafa}(1989)}]{Brandenberger:1988:sit}
\bibinfo{author}{\bibfnamefont{R.~H.} \bibnamefont{Brandenberger}}
  \bibnamefont{and} \bibinfo{author}{\bibfnamefont{C.}~\bibnamefont{Vafa}},
  \bibinfo{journal}{Nucl. Phys.} \textbf{\bibinfo{volume}{B316}},
  \bibinfo{pages}{391} (\bibinfo{year}{1989}).

\bibitem[{\citenamefont{Randall and Sundrum}(1999)}]{randall:1999:atc}
\bibinfo{author}{\bibfnamefont{L.}~\bibnamefont{Randall}} \bibnamefont{and}
  \bibinfo{author}{\bibfnamefont{R.}~\bibnamefont{Sundrum}},
  \bibinfo{journal}{Phys. Rev. Lett.} \textbf{\bibinfo{volume}{83}},
  \bibinfo{pages}{4690} (\bibinfo{year}{1999}), \eprint{hep-th/9906064}.

\bibitem{Alexander:2007vt}
  S.~Alexander and N.~Yunes,
  Phys.\ Rev.\  D {\bf 75}, 124022 (2007)
  [arXiv:0704.0299 [hep-th]].

\bibitem[{\citenamefont{Blanchet}(2006)}]{Blanchet:2002av}
\bibinfo{author}{\bibfnamefont{L.}~\bibnamefont{Blanchet}},
  \bibinfo{journal}{Living Rev. Rel.} \textbf{\bibinfo{volume}{9}},
  \bibinfo{pages}{4} (\bibinfo{year}{2006}), \bibinfo{note}{and references
  therein}, \eprint{gr-qc/0202016}.

\bibitem[{\citenamefont{Alexander et~al.}(2007)\citenamefont{Alexander, Finn,
  and Yunes}}]{Alexander:2007:bgw}
\bibinfo{author}{\bibfnamefont{S.}~\bibnamefont{Alexander}},
  \bibinfo{author}{\bibfnamefont{L.~S.} \bibnamefont{Finn}}, \bibnamefont{and}
  \bibinfo{author}{\bibfnamefont{N.}~\bibnamefont{Yunes}}, \bibinfo{journal}{in
  progress}  (\bibinfo{year}{2007}).

\bibitem{Will:2002ma}
  C.~M.~Will,
  Phys.\ Rev.\  D {\bf 67}, 062003 (2003)
  [arXiv:gr-qc/0212069].

\bibitem{Smith:2007jm}
  T.~L.~Smith, A.~L.~Erickcek, R.~R.~Caldwell and M.~Kamionkowski,
  arXiv:0708.0001 [astro-ph].

\end{thebibliography}
\end{document}